\documentclass[reprint,amsmath,amssymb,apr,aip,onecolumn]{revtex4-2}
\usepackage{graphicx}
\usepackage{graphics}
\usepackage{algorithm,algorithmic}
\usepackage{mathptmx}
\usepackage{times}
\usepackage{amsmath}
\usepackage{amssymb}
\usepackage{dcolumn}
\usepackage{color}
\usepackage{bm}
\usepackage{units}
\usepackage{epsfig}
\usepackage{soul}
\usepackage{subfiles}

\newcommand{\ignore}[1]{}

\begin{document}

\title{Rapid yet accurate Tile-circuit and device modeling for Analog In-Memory\\ Computing}

\author{
J. Luquin$^1$,            
C. Mackin$^1$,            
S. Ambrogio$^1$,          
A. Chen$^1$,              
F. Baldi$^1$,             
G. Miralles$^1$,           
M. J. Rasch$^2$,          
J. Büchel$^3$,            
\\
M. Lalwani$^1$,           
W. Ponghiran$^2$,         
P. Solomon$^2$,           
H. Tsai$^1$,              
G. W. Burr$^1$,          
and P. Narayanan$^1$      
}
\affiliation{IBM Research -- Almaden, 650 Harry Road, San Jose, CA~~USA.}
\affiliation{IBM T. J. Watson Research Center -- Yorktown Heights, NY~10598, USA.}
\affiliation{IBM Research –- Zurich, Rüschlikon, Switzerland.}

\maketitle

\textbf{Abstract}

Analog In-Memory Compute (AIMC) can improve the energy efficiency of Deep Learning by orders of magnitude.
Yet analog-domain device and circuit non-idealities -- within the analog ``Tiles'' performing Matrix-Vector Multiply (MVM) operations -- can degrade neural-network task accuracy. 
We quantify the impact of low-level distortions and noise, and develop a mathematical model for Multiply-ACcumulate (MAC) operations mapped to analog tiles. 
Instantaneous-current IR-drop (the most significant circuit non-ideality), and ADC quantization effects are fully captured by this model, which can predict MVM tile-outputs both rapidly and accurately, as compared to much slower rigorous circuit simulations. 
A statistical model of PCM read noise at nanosecond timescales is derived from -- and matched against -- experimental measurements. 
We integrate these (statistical) device and (deterministic) circuit effects into a PyTorch-based framework to assess the accuracy impact on the BERT and ALBERT Transformer networks. 
We show that hardware-aware fine-tuning using simple Gaussian noise provides resilience against ADC quantization and PCM read noise effects, but is less effective against IR-drop.
This is because IR-drop -- although deterministic -- is non-linear, is changing significantly during the time-integration window, and is ultimately dependent on all the excitations being introduced in parallel into the analog tile. 
The apparent inability of simple Gaussian noise applied during training to properly prepare a DNN network for IR-drop during inference implies that more complex training approaches -- incorporating advances such as the Tile-circuit model introduced here -- 
will be critical for resilient deployment of large neural networks onto AIMC hardware.

\section{Introduction}\label{sec:intro}

Deep Neural Networks (DNNs) have demonstrated unparalleled capabilities in recent years for  applications such as image processing, natural language understanding, and content generation. 
However, much of this progress has been driven by larger DNN-model sizes, which has exponentially increased compute, memory, and energy requirements. 
This growing computational demand greatly incentivizes  hardware solutions that can reduce the energy consumed while training and using (performing ``inference'' with) large DNNs.

Analog in-memory computing (AIMC) is a promising option for DNN hardware acceleration.
AIMC ``Tiles'' implement the large Multiply-ACcumulate (MAC) operations that dominate DNNs in a massively parallel fashion without the need for data movement, leading to increased throughput and energy efficiency~\cite{architecture_tvlsi, peng2020dnn+, shafiee:2016}. 
A capacitive AIMC uses current and time for multiplication, storing the results on a capacitive array, followed by charge sharing between capacitors for accumulation. 
A resistive AIMC generates individual currents as the multiplication of a (time-encoded) voltage signal and a conductance, and accumulates them per Kirchhoff's current law. 
In either case, AIMC Tiles combine the underlying physics and topology of capacitor- or resistor-arrays to implement MACs efficiently. 
Variations to these general schemes enable signed activations, signed weights, and multi-bit computation.

The benefits of AIMC do not come without challenges. 
Device and circuit non-idealities can cause deviations from the expected or ideal MAC output. 
For instance, resistive AIMC uses non-volatile memories (NVM) such as Resistive RAM (RRAM) and Phase Change Memory (PCM), which exhibit varying extents of programming error, read noise and resistance drift -- implying that the analog conductances used in the multiply operation are inherently imprecise. 
Finite wire and driver resistance along the accumulation paths leads to instantaneous-current IR-drop distortions, which depend both on the locations of the inputs being activated, and on the values of the conductances at those inputs. 
In time-based encoding schemes, the IR-drop also changes at each time-step as activation patterns change. 
Peripheral circuitry that converts the analog MAC into a digital quantity for downstream DNN processing (typically a sense stage followed by an analog-to-digital converter (ADC)) can also be non-ideal, with variability in offset, slope and non-linear characteristics introducing yet more error.

Fortunately, DNNs are resilient to imprecise computation in many cases, as evidenced by the large body of work on reduced precision models that achieve benefits in DNN-model size, throughput, and energy efficiency without sacrificing accuracy~\cite{nagel2020up, liu2021post, wang2022deep, frantar2022gptq}. 
Weight and activation quantization is achieved using variants of post training quantization (PTQ) or quantization-aware training (QAT) techniques, which scale network hyper-parameters and/or fine-tune the DNN-model through additional training epochs, exposing the forward pass to low precision compute and allowing  backpropagation to correct for the `noise' introduced. 
Mirroring this approach, hardware-aware (HWA) training~\cite{raschHWA:2023} exposes the forward pass to simulated AIMC hardware distortions, ideally recovering all the accuracy loss from deterministic analog imperfections, while also making the network highly resilient to random analog noise.

While many hardware imperfections -- such as NVM read noise and ADC noise -- are independent random noise sources that can be modeled using well-defined probability distributions, instantaneous-current IR-drop is an important exception.  
In addition to the time- and position-dependence described earlier, IR-drop tends to compound downstream circuit imprecision. 
That said, IR-drop for a given weight-matrix and activation-vector is completely deterministic. 
This implies that, given an analytical approach -- a ``model'' -- for predicting the effects of IR-drop with sufficient fidelity, one may be able to completely capture its impact in inference simulations, and if necessary, fine-tune networks to help counteract it through HWA training.  
The latter would also require that the activation-vectors encountered during training strongly resemble those encountered during inference.
Even without retraining, an accurate and comprehensive model of the physical hardware incorporating both random and deterministic sources is tremendously useful, making it feasible to gauge the relative impacts of subtle, low-level behaviors on overall DNN accuracy of large networks, something that is difficult even with physical hardware.
That said, to be widely used -- either in improving training or in assessing inference -- this predictive analog Tile-circuit model needs to be computationally inexpensive yet offer nearly the same high-fidelity as full circuit simulations.

The primary contribution of this article is to develop such a model. 
We use detailed circuit-level simulations, in conjunction with realistic device characterization data at MAC-relevant time scales, to quantify the error in the MAC operation. 
This simulation framework captures imperfections in the crossbar arrays, including resistive and capacitive parasitics, operational amplifier and ADC non-idealities, and device read noise. 
The analysis is done under various circuit conditions to identify configurations that are advantageous in an energy/accuracy trade-off landscape. 
We then build simplified models of the MAC incorporating the non-idealities, showing near identical correspondence to circuit simulation. 
We integrate these hardware non-ideality models into IBM's open source AI-Hardware toolkit~\cite{aihwkit_aicas, le2023using}, quantifying the impact on large DNNs. 
Finally, we demonstrate that hardware-aware fine-tuning based on the analytical models of the MAC can recover lost accuracy, making the DNN models resilient to real analog device and circuit distortions.

\begin{figure*}[t]
\centering
\centerline{\epsfig{file=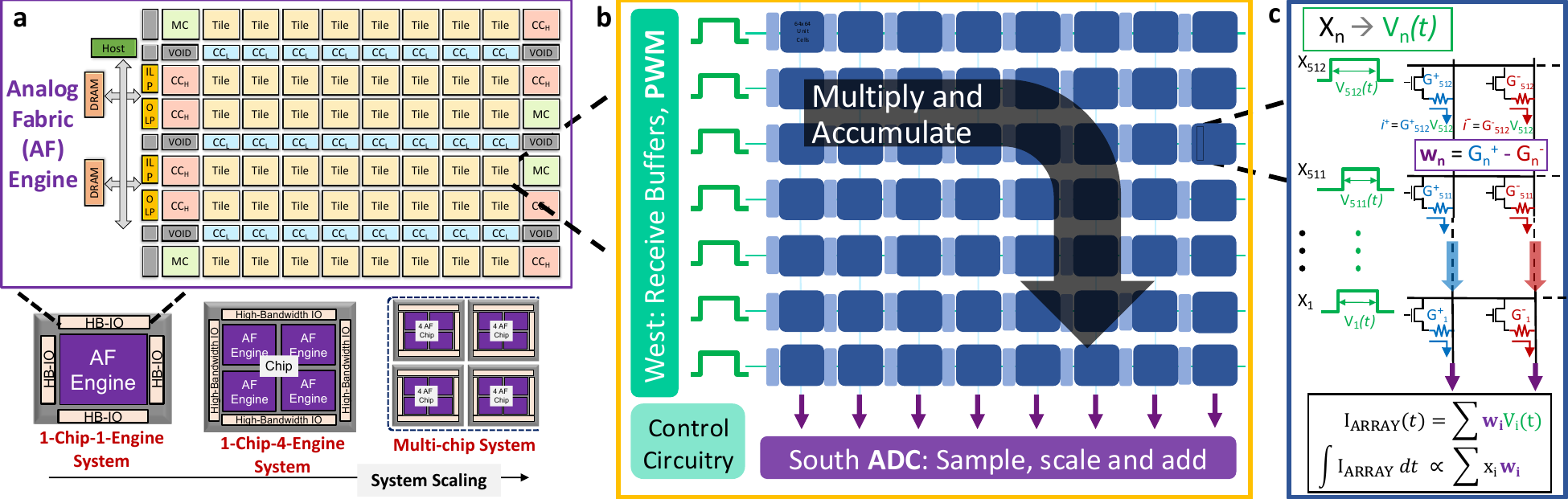, width= 1\columnwidth}}
\caption{\textbf{AIMC Chip Overview} depicting (a) the system level architecture of an ``Analog Fabric Engine'' for mapping Deep Neural Networks onto a 2-D array of heterogenous compute resources. 
MAC operations are performed at the location of weights programmed into analog Tiles. 
``Heavy'' and ``Light'' compute cores (CC$_\mathrm{H}$ and CC$_\mathrm{L}$), along with SRAM scratchpads in Memory Cores (MC), implement auxiliary neural network digital operations. 
A 2-D runtime-configurable mesh allows just-in-time vector transport across the chip. 
AFEs can be tiled together to scale up to systems that map large neural networks. 
(b) Each Tile consists of a 512$\times$512 array of Unit Cells (UC). 
Activation signals are injected into the tiles through West Periphery Circuitry. 
Currents generated at crosspoints are accumulated along columns and connected to south ADCs, where results of the MAC operations are processed, integrated, and converted to digital numbers. 
(c) The multiply operation is performed with the generation of a current proportional to $G \cdot V$ in each unit cell (Ohm’s Law), while accumulation is performed using Kirchhoff’s Law to sum all the currents flowing south along the vertical bitline. 
Signed weights are implemented using two conductances, $G$+ and $G$--, allowing bidirectional individual and cumulative currents depending on the signs of the weights and activations. 
}
\label{fig:overview}
\vspace{-0.15in}
\end{figure*}

\section{Analog In-Memory Multiply-Accumulate (MAC) Overview}\label{sec:overview}

Fig.~\ref{fig:overview}a shows the system-level architecture of an AIMC chip, comprised of Analog Fabric (AF) Engines and an Input/Output (I/O) interface. 
To map an end-to-end neural network, each AF consists of a number of analog Tiles that implement the multiply-accumulate operations, together with nearby digital Compute Cores for auxiliary operations such as activation or normalization operations needed by the DNN. 
Data transfer between components within an AF is achieved using a highly parallel 2-D Mesh over the cores and tiles~\cite{architecture_tvlsi, ares_nature}. 
Multiple such engines and chips can be combined together into larger systems that can map large neural networks. 
Functional verification (and if needed, hardware-aware training) of such systems requires accurate modeling of the individual components. 
Fortunately, data transfer can be made error free, and digital operations can be modeled in a straightforward fashion, including the effects of clipping and rounding depending on local compute precision.
However, modeling the behavior of the analog Tiles is decidedly non-trivial.

Fig.~\ref{fig:overview}b zooms into the structure of a single analog Tile. 
On the West edge of each tile is periphery circuitry to buffer and direct activation signals into the Tile; on the South edge, in addition to peripheral control circuitry, per-column ADCs are responsible for integrating currents along the column, performing the ACcumulate operation. 
The tile consists of a 512$\times$512 array of Unit Cells (UC), where each UC performs a single Multiply operation; an abstracted view of a single logical column, showing Unit Cells and MAC operation, is illustrated in Fig.~\ref{fig:overview}c.

\begin{figure*}[t]
\centering
\centerline{\epsfig{file=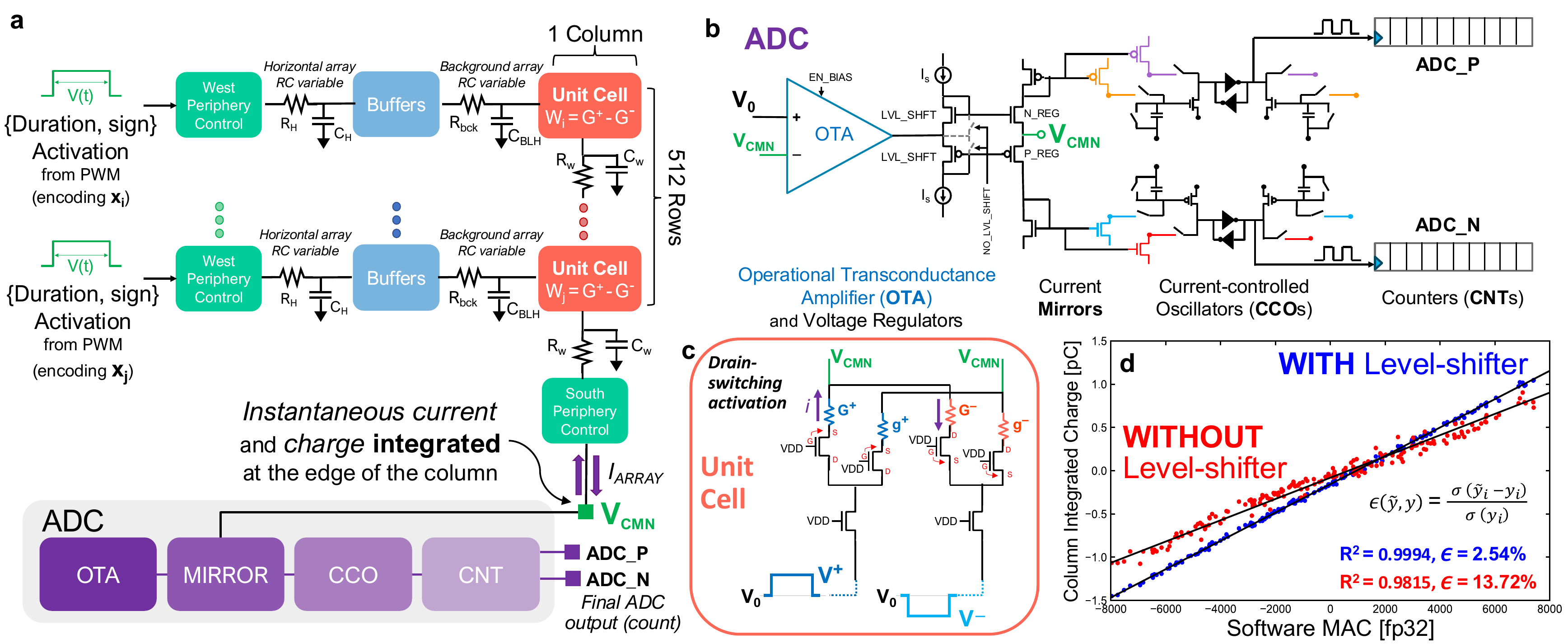, width= 1\columnwidth}}
\caption{\textbf{Circuit simulation framework generates ground-truth MAC data to support model generation} using (a) a single column Tile schematic. 
Activation signals are buffered into the unit cells through the West circuitry. 
NVM elements within unit-cells can include arbitrary VerilogA-based noise models. 
Parasitics are distributed throughout rows and the column. 
Current flowing through the column is fed into the South ADC circuitry, comprising an OTA and a pair of current mirrors, current controlled oscillators (CCO) and digital counters (CNT) for the analog to digital conversion. 
(b) Circuit schematic of the analog-to-digital converter (ADC) where the column current is mirrored and fed into the CCO. 
The OTA is used as an analog buffer to keep the read voltage stable at the desired level of $V_0=\:$400mV as array impedance changes. 
An optional current level-shifter circuit keeps Class AB regulators out of deep subthreshold operation at low array current. 
Each CCO generates voltage pulses at a frequency proportional to its portion of the input current.
With the level-shifter OFF, typically only one CCO is oscillating; with the level-shifter ON, both CCOs may be active. 
The pulses are counted using digital counters and stored in registers to generate a digital output by subtraction after simple gain and offset correction (not shown).
Calibration bits (not shown) serve to adjust the current mirror gains and to adjust feedback\cite{khaddam2021hermes} to enforce a linear frequency-vs-current response.
(c) In the drain-switching activation scheme, activation pulses are driven to the drain terminal of the lower select transistors. 
The positive or negative sign of the activation determines the polarity of the differential read voltage applied to the conductance pairs, with the opposite terminal held at the common-mode voltage by the OTA sitting at the end of the vertical bitline.
Note that PCM programming constraints ($V_{PCM} \gg$ max-allowed single-transistor $V_{DS}$) leads to a unit-cell with two stacked transistors.  
(d) Simulated MAC accuracy based on circuit simulations compared to ideal software MAC values are shown, with and without the level-shifter circuitry enabled.
}
\label{fig:simulationFramework}
\vspace{-0.15in}
\end{figure*}

To implement the MAC operation, a vector of activations is applied to all 512 rows in a parallel fashion. 
In this study, the activation magnitude is encoded into the duration of fixed-amplitude voltage pulses. 
The sign of each activation sets the `direction' of the voltage potential across the resistive device. 
As illustrated in Fig.~\ref{fig:simulationFramework}c, one end of the resistive device is tied to a common voltage, $V_{cmn}$, held at $V_0 =$ 400mV $= V_{DD}/2$, while the other end is tied to either $V^{+}$ or $V^{-}$, at 600mV or 200mV, respectively. 
As this activation signal control is accomplished with individual row-wise periphery circuitry on the West edge of the tile, this `differential' encoding scheme allows a single inference pass to integrate both positive and negative activations and to act on signed weights programmed onto a PCM conductance pair, $G$+ and $G$--. 
By allowing currents to be added or subtracted within the array, this approach also reduces IR-drop along the column. 
The net current from all rows over the total integration time is integrated at a per-column current-based ADC\cite{khaddam2021hermes}. 

Instantaneous current from each column flows into or out of the ADC~\cite{khaddam2021hermes, hermes2023}, shown in Fig.~\ref{fig:simulationFramework}b. 
An Operational Transconductance Amplifier (OTA) is responsible for maintaining $V_{cmn}$ nominally at $V_0=\:$400mV. 
The read current then enters two current mirrors, whose calibration parameters are tunable on a column-wise basis and to adjust the gain of the ADC. 
One current mirror is dedicated to currents flowing into the ADC from the array (``positive'' current); the other to currents flowing from the ADC into the array (``negative'' current).
The mirrored currents each charge a  capacitor in the appropriate Current Controlled Oscillator (CCO). 
Once one of these capacitors reaches a fully-charged state, it trips a latch in the CCO which switches integration to a second capacitor, which then trips the latch again after charge accumulation and so forth, generating oscillations. 
The frequency of oscillation is dependent on the current from the mirror (and thus, the current from the array). 
Each oscillation generates a pulse, which is then counted using digital counters in the final stage of the ADC. 
The final, signed-integer MAC value is produced by subtracting the ``negative'' counter value from the ``positive'' counter value, with a simple digital correction applied to compensate for gain and offset differences between the two current mirrors (not shown).
The digital portions of the ADC use $V_{DD}=\:$800mV.

\section{Circuit Simulation Framework}\label{sec:colSim}

A baseline circuit simulation framework (Fig.~\ref{fig:simulationFramework}a) is used to generate ground-truth data to support the development of the analog Tile-circuit model developed in the next section. 
To explore the accuracy of Analog MAC operations with reasonable computational resources, the simulation framework comprises just a single column (a 512$\times$1 Tile), albeit in full detail including all row drivers, the column parasitics and one full ADC, abstracting the row-wise elements of the array into lumped components. 
Tuning row-wise RC parameters changes the 'virtual' location of the column to be any of the 512 columns in the tile, impacting the signal integrity of the row-wise input signals. 
The column parasitics are distributed across the entire column, fully capturing voltage degradation effects due to instantaneous-current IR-drop, as well as transient impacts such as activation changes within an integration period. 

The simulations are conducted using Spectre and 14nm foundry models for the transistors. 
Custom VerilogA models are used for the PCM, incorporating read noise (described in Section~\ref{sec:deviceModel}) but the methodology is similarly applicable to any NVM. 
Both the weight- and activation-vectors are randomly generated from Gaussian distributions; the range of weights and activations allow the array to generate current ranging from -100uA to 100uA. 
That said, the simulation framework can accommodate any distribution of input data including real activations and weights from trained DNN models. 
This flexibility allows for thorough analysis regarding the behavior of the circuitry under various operating conditions. 

With the aforementioned conditions, two 200-point sweeps of weight and activation input vectors were simulated to obtain baseline MAC results.
Fig.~\ref{fig:simulationFramework}d shows a scatter plot of actual (simulated) vs.\ expected MAC under two different ADC operating conditions.
Each datapoint thus represents the simulated MAC result for a full sum across 512 activations, $x_i$, multiplying 512 weights, $w_i$, followed by the summation of these 512 $x_i\:w_i$ product-terms.
When the current level shift circuits shown in Fig.~\ref{fig:simulationFramework}b are disabled, the error is higher, since the class AB regulators maintaining the read voltage end up operating in the deep subthreshold regime whenever the array current-magnitude is low (see Supplementary Materials). 
However, when the level-shifter is enabled, its injected bias current helps keep the regulated voltage stable across the full range of array currents. 
All subsequent results will be shown with the level-shifter enabled.

In practical use, the output of each ADC will be processed by a vectorized affine scale and offset\cite{raschHWA:2023}, which among other tasks, can help correct the overall ADC gain and offset, and make all ADCs appear identical to the system.
Thus it is important that all of the datapoints fall on the same line, with the actual slope and offset of this line being relatively unimportant (within some reasonable range). 
In order to quantify deviations from this linear fit-line, our error metric, $\epsilon(\tilde{y}, y)$,
is defined as the ratio of the standard deviation of the error to the standard deviation of the desired output, as shown in the inset of Fig.~\ref{fig:simulationFramework}d. 
This metric captures the extent of the error relative to the signal in a mean- and offset-invariant manner, and will be used in the rest of the paper to quantify correlations between various quantities.

\section{Python-based Tile Circuit Model}\label{sec:circuitModel}

\begin{figure}[t]
\centering
\centerline{\epsfig{file=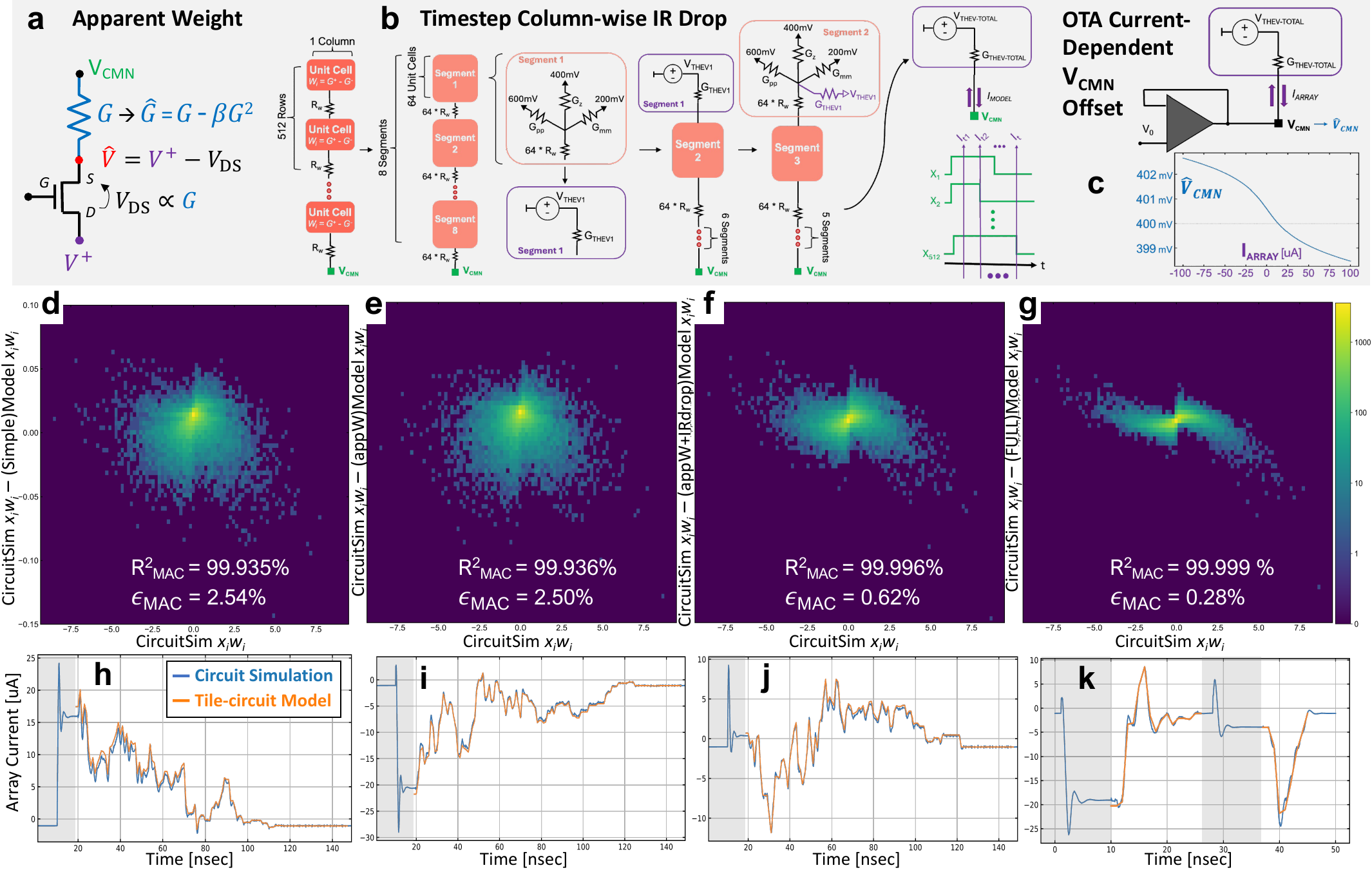, width=0.9\columnwidth}}
\caption{\textbf{Circuit non-ideality models} include an 
(a) ``apparent weight'' that arises from the non-ideal switching behavior of the select transistor, caused by the small but non-zero $V_{DS}$ across its source and drain terminals. 
Since $V_{DS}$ is proportional to the $I_{DS}$ current, and the $I_{DS}$ current is proportional to the programmed $G$, this phenomenon reduces the effective weight of the cell as captured by a simple quadratic expression. 
(b) To efficiently estimate the impact of instantaneous-current IR-drop, the parasitic resistance distributed across 512 unit cells in every column is first lumped into groups of 64 elements (with the unit cells in parallel), and then reduced to a Thevenin equivalent model. 
Iterative Thevenin reductions produce a simplified equivalent circuit snapshot of the column for a given set of active inputs. 
As the landscape of activations changes, the Thevenin equivalent must be recalculated at each discrete (in this case 1 ns) time-step. 
(c) Weak current dependence in the OTA buffer can cause the read voltage to differ from the ideal value of $V_0=\:$400mV as a function of the current flowing in/out of the ADC block. 
(d)-(g) Error heat maps comparing the error between $x_i\:w_i$ product-term as predicted by our Tile-circuit model and the results of rigorous but slow circuit simulations, as a function of the $x_i\:w_i$ product-term.  
Ideally these would be horizontal lines, showing zero discrepancy between the model and circuit-simulations across all possible product term values. 
Annotations show the R$^2$ coefficient and $\epsilon$ values for the resulting MAC (the sum across all product terms) for correlation plots that resemble~Fig.~\ref{fig:simulationFramework}d, except plotting model prediction as a function of circuit simulation ground-truth. 
R$^2_\mathrm{MAC}$~($\epsilon_\mathrm{MAC}$) reaching 100\%~(0.0) would imply that the Tile-circuit model exactly matches the circuit simulations.
(d) shows a simple linear version of the model, with (e) adding the apparent weight due to transistor unit-cell $V_{DS}$, (f) further adding the instantaneous-current IR-drop, and (g) further including OTA imperfections. 
Clearly, adding the IR-drop (from (e) to (f)) has the biggest impact on the fidelity of the Tile-circuit model. 
(h)-(k) Results are shown for four scenarios, three using conventional PWM excitation (127ns for 7 amplitude-bits, (h,i,j)), one using a hybrid Split-mode PWM excitation (two pulses, 15ns for top 4 bits, then 7ns for bottom 3 bits, see Section~\ref{sec:splitPWM}). 
In (h), the MAC is large and positive; in (i), the MAC is large but negative; in both (j) and (k), the current changes direction during the integration period(s). 
In all four cases, excellent correspondence is observed between transient results of Python-based Tile-circuit model and Spectre circuit simulations.
}
\label{fig:circuitModel}
\vspace{-0.15in}
\end{figure}

In this section, we develop a python-based Tile-circuit model to abstract circuit behaviors to capture the MAC error in Fig.~\ref{fig:simulationFramework}d across a range of input and weight vector conditions without running full circuit simulations. 
Creating a lightweight model is critical for integration of low-level circuit behavior into high-level Deep Learning frameworks such as PyTorch and to extending MAC accuracy studies to large DNN models.

\subsection{Apparent Weight}\label{apparent_weight}

The ``apparent weight'' effect is illustrated in Fig.~\ref{fig:circuitModel}a. 
In the ideal case, each PCM device in a Unit Cell is tied to $V_{cmn}$ on one side and -- through a pair of stacked select-transistors -- $V^{+}$/$V^{-}$ on the other, creating a $V_{read}$ potential across the device.  In this case, $|V_{read}|$ is ideally 200mV. 
However, when the select transistor is ON, a small but non-zero $V_{DS}$ appears across the FET's Drain/Source terminals, modifying the voltage at the upstream terminal of the PCM device from $V^{+}$ (or $V^{-}$) to $\hat{V}$. 
Since this voltage drop across the FET is a function of $I_{DS}$, which in turn is a function of the programmed conductance on the PCM device, $\hat{V}$ depends linearly on the magnitude of the programmed conductance, $G$. 
The effective conductance $\hat{G}$, otherwise referred to as the apparent weight, can then be calculated using a simple quadratic function of $G$, using a $\beta$ coefficient obtained from circuit simulations of the select transistor(s).

\subsection{IR-Drop}\label{IR_drop}

As will be demonstrated soon, the primary component of the residual MAC error comes from the IR-drop. 
In circuit simulations, wire resistances were explicitly included with values obtained through post-layout parasitic extraction. 
As shown in Fig.~\ref{fig:circuitModel}b, ADCs maintain a common voltage, $V_{cmn}$, at the bottom of each column. 
However, the current flow in each column will incur IR-Drop and cause deviations on the actual read voltage across UCs that not only vary from column to column and within each column, but also depend on the instantaneous inputs (how many rows are activated) at that instant in time. 
Including IR-drop along a column is vital to modeling the transient behavior of the column current, and thus to calculating an accurate MAC output value. 
However, running full circuit-simulations across 512 rows can be prohibitively slow, so approximation is essential.

Fig.~\ref{fig:circuitModel}b illustrates the approximate Tile-circuit model we have developed, which clumps 64 unit cells into one segment, and the wire parasitics into one single lumped resistor. 
Since the unit cells are now parallel, their conductances can be added together into three lumped conductors, one connecting to the positive differential voltage ($V^+$) and one to the negative differential voltage ($V^{-}$), covering activations that are non-zero at a given time-step, and one to the common voltage ($V_0$), covering activations that are zero. 
As shown in Fig.~\ref{fig:circuitModel}b, these can then be simplified by Thevenin-equivalent reduction to a single voltage and resistance. 
Once the top segment in reduced, the process can be repeated across all subsequent segments, collapsing the column down to a single equivalent conductance and voltage from which an instantaneous current can be calculated. 
The process can be repeated in each subsequent time-step as various activation pulses shut off (or start up). 
Ultimately, the python Tile-circuit model takes software weights and activations as the input to perform this IR-drop abstraction process and accordingly produce a final MAC estimation.

\subsection{OTA voltage deviations}\label{OTA_voltage}

Up to now, $V_{cmn}$ has been referred to as a consistently held, common node voltage important for setting the read voltage across the PCM device during a MAC operation, nominally set to $V_0=\:$400mV. 
However, one final non-ideality explored via the Tile-circuit model analysis is the deviation of $V_{cmn}$ from its ideal value at the OTA (Operational Transconductance Amplifier). 
The OTA in the ADC is responsible for maintaining $V_{cmn}$ at $V_0$, irrespective of the current generated by the column. 
Fig.~\ref{fig:circuitModel}c illustrates that the presence of instantaneous current from the array column can cause the output voltage of the OTA to deviate by a few milli-Volts from the target value of $V_0$. 
Our Tile-circuit model incorporates this non-ideal OTA behavior via a lookup table -- representing rigorous circuit simulation of the OTA circuit -- to capture the residual MAC errors.

\subsection{Tile-circuit model as compared to full circuit simulations}\label{model_comparison}

We can understand the relative impact of the non-idealities described above 
by quantifying the error -- between individual $x_i\:w_i$ product terms as predicted by variants of our Tile-circuit model, 
and the precise $x_i\:w_i$ terms obtained from circuit simulations -- as we make our Tile-circuit model increasingly more realistic by incorporating more non-idealities within it. 
Across the 200 sampled MACs mentioned earlier, we have 102,400 raw $x_i\:w_i$ product terms we can look at. 
We plot heat maps of error between model-variant and circuit-simulations as a function of circuit-simulated ground-truth $x_i\:w_i$ in Fig.~\ref{fig:circuitModel}d-g. Ideally, the error would be identically zero across all $x_i\:w_i$ magnitudes, and the annotated R$^2_\mathrm{MAC}$ and $\epsilon_\mathrm{MAC}$ values would reach 100\% and 0.0, respectively, indicating that not just the individual product-terms, but the resulting summed MAC values, were indistinguishable between the Tile-circuit model and rigorous circuit simulation.

Fig.~\ref{fig:circuitModel}d shows heat maps for a simple linear model incorporating none of our specific non-idealities, and then apparent weight, IR-drop, and OTA voltage offsets are added in sequence in Figs.~\ref{fig:circuitModel}e, f, and g. 
The apparent weight effect does not markedly change the error heat-map and produces only a small improvement in R$^2_\mathrm{MAC}$ and $\epsilon_\mathrm{MAC}$. 
However, once the instantaneous-current IR-drop is included in our Tile-circuit model (Fig.~\ref{fig:circuitModel}f), the error heat-map, R$^2_\mathrm{MAC}$ and $\epsilon_\mathrm{MAC}$ all
show significant improvement, indicating that IR-drop is the most important effect that needs to be captured. 
Finally, incorporating OTA non-ideality (Fig.~\ref{fig:circuitModel}f) provides a small but non-negligible further improvement in the ability of our python-based Tile-circuit model to predict the same product-terms and summed MAC results as the full circuit simulations.

Fig.~\ref{fig:circuitModel}h-k provide another visualization of the accuracy of the developed python-based Tile-circuit model, overlaying the transient current waveforms produced by the circuit simulation (blue) with those produced by our Tile-circuit model (orange), for four representative activation-vector scenarios.  These include scenarios where weight-terms are: consistently positive leading to a large positive MAC 
(Fig.~\ref{fig:circuitModel}h); consistently negative leading to a large-magnitude negative MAC result (Fig.~\ref{fig:circuitModel}i); 
change sign during the integration so that large instantaneous currents subtract to produce a small MAC result (Fig.~\ref{fig:circuitModel}j); or are summed in two separate intervals using the ``Split'' PWM mode described below (Section~\ref{sec:splitPWM}, Fig.~\ref{fig:circuitModel}k).
Collectively, these results demonstrate that our Tile-circuit model performs equally well across different current polarities, input activation modes, and activation trajectories.

\subsection{PCM Device Model}\label{sec:deviceModel}

\begin{figure}[t]
\centering
\centerline{\epsfig{file=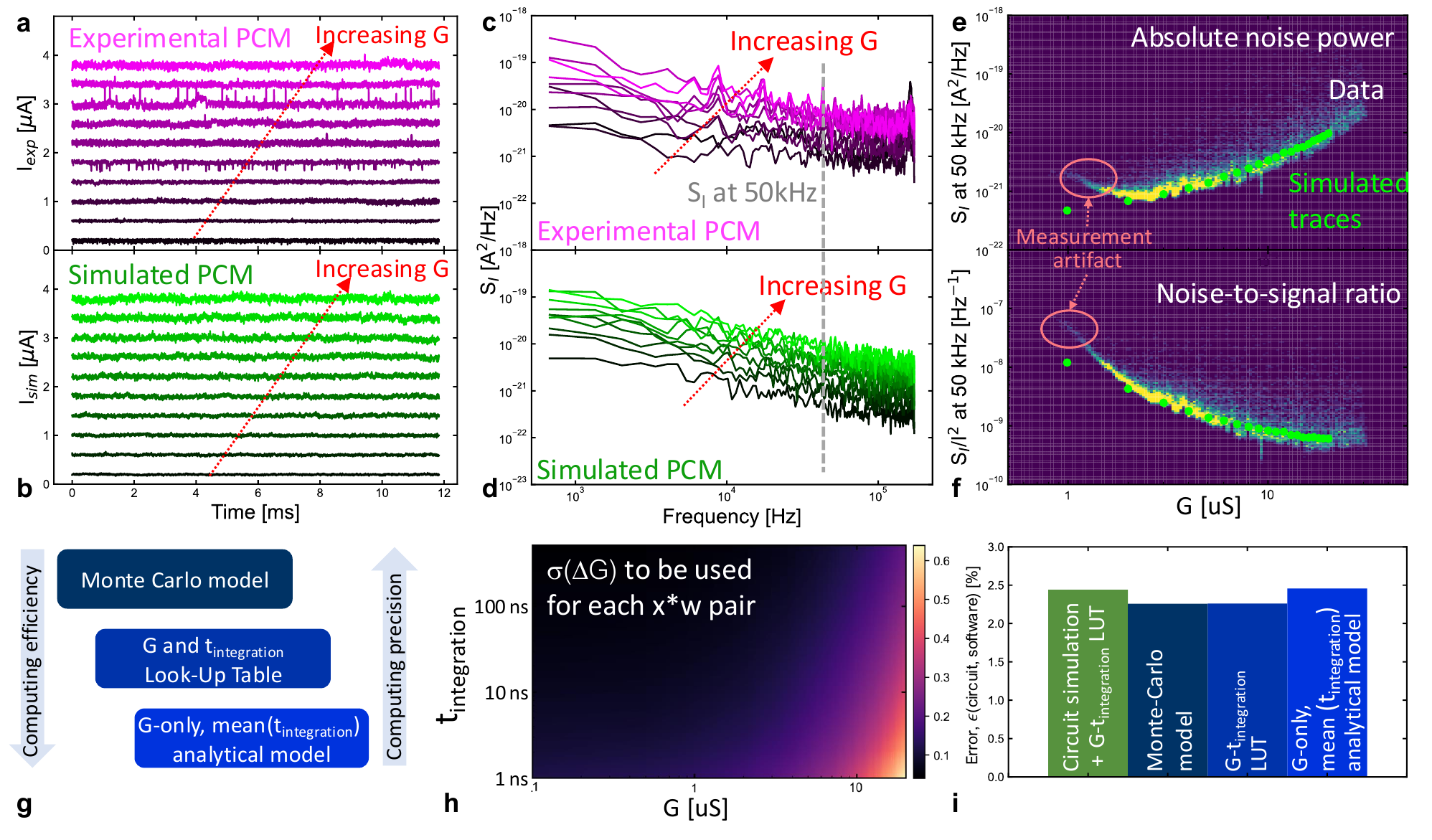, width=1\columnwidth}}
\caption{\textbf{PCM Noise Model}: (a) Experimental current-traces from PCM devices with a 200mV read voltage at millisecond time scales. (b) Simulated PCM current-traces using Monte-Carlo based noise-model. 
(c)-(d) Respective power spectra for (a)-(b) showing a -1 slope, characteristic of 1/f noise behavior. 
(e) Absolute noise power and (f) noise-to-signal ratio, both at 50 kHz, of experimental data as a function of PCM device-conductance; simulation data from our PCM noise-model is overlaid on the same plots. 
Experimental data exhibit a measurement artifact at low conductances due to high-frequency measurement-noise, 
which is not modeled and not present in the overlaid simulated data. 
(g) PCM device noise can be modeled with various levels of abstraction/complexity, where increasing precision of the noise-model comes at a computational efficiency cost. 
(h) At one level of abstraction, the PCM noise behavior can be modeled as a Lookup Table (LUT), where a Gaussian-noise effective sigma can be extracted for a PCM device dependent on the device conductance ($G$, proportional to the programmed weight, $w_i$) and the device current integration time ($t_{integration}$, 
proportional to the DNN input, $x_i$). 
(i) MAC errors obtained from circuit simulations incorporating the LUT PCM noise-model into our python-based Tile-circuit model, using a variety of PCM read noise-models, demonstrate comparable MAC performance, showing that approximate (and computationally-efficient) versions of our PCM noise-model can offer high fidelity predictions.
}
\label{fig:pcmModel}
\vspace{-0.15in}
\end{figure}

In addition to circuit non-idealities, read noise in non-volatile memory devices can have a strong time-dependent component that needs to be captured. 
Many types of PCM noise~\cite{papandreou2011programming, nandakumar2019phase, ambrogio:2019}, such as programming noise, read noise, device-to-device variability, and drift, and their effects on DNN accuracy have been modeled using DNN frameworks~\cite{mackin:2019, joshi:2020_zurich, raschHWA:2023, le2023using}. 
Since this read noise is `upstream' from the circuits and can combine with IR-drop and ADC quantization error in non-linear ways, in this work  we quantify the dependence of PCM device read-noise on both programmed device-conductance $G$ (and thus weight $w_i$) and device read-duration $t_{integration}$ (and thus DNN input $x_i$) and capture this in our analytical Tile-circuit model. The general methodology is applicable to other NVM device-types beyond PCM.

We first conducted experimental measurements of PCM devices to gather statistical data. 
More details on the experimental setup for acquiring PCM device data are provided in the Supplementary Materials. 
Fig.~\ref{fig:pcmModel}a and Fig.~\ref{fig:pcmModel}c show example current traces from PCM devices and their corresponding power spectra. 
Since the power spectra are dominated by $1/f$ noise, we constructed a Monte-Carlo simulation using this $1/f$ noise approximation. 
Fig.~\ref{fig:pcmModel}b and Fig.~\ref{fig:pcmModel}d shows that simulated current-traces and noise-spectra closely resemble the measured experimental data. 
We observe consistent trends between the absolute noise power and noise-to-signal ratio as a function of PCM conductance at a specific frequency (here 50kHz). 
Fig.~\ref{fig:pcmModel}e-f shows that PCM devices with higher conductances have higher absolute noise, but lower noise-to-signal ratios. 
Monte Carlo simulation results overlay nicely onto the experimental data, illustrating that this PCM noise-model is accurate. 

While the Monte Carlo simulations accurately track the experimental data, the resulting noise-model doesn't run fast enough to be feasibly used in a high-level DNN framework. 
Fig.~\ref{fig:pcmModel}g outlines a series of model abstractions to obtain a balanced trade-off between compute efficiency and model precision. 
The first abstraction is to use the Monte Carlo (MC) noise-model to simulate $\Delta G$, the difference between integrated conductance under read noise and the nominal conductance. 
Over multiple MC samples of device conductance at different integration times, the standard deviation in ${\Delta}G$ can be captured in a Lookup Table (LUT) as shown in Fig.~\ref{fig:pcmModel}h.  Fig.~\ref{fig:pcmModel}h shows that integrating higher conductances for shorter periods of time results in higher read noise. 
Once a given $x_i$ and $w_i$ value are identified, the resulting $t_{integration}$ and $G$ are used to obtain the effective Gaussian sigma for the $x_i\:w_i$ term from 
Fig.~\ref{fig:pcmModel}h.

While it is possible to use the full LUT and sample each conductance independently based on the input in the python-based framework too, a second level of abstraction for efficiency determines the standard deviation as a function of the mean activation magnitude for a given batch, instead of looking up $\sigma(\Delta G)$ for each individual $x_i\:w_i$ term.  Fig.~\ref{fig:pcmModel}i shows that this abstracted PCM-noise models produce comparable MAC errors to the LUT model, and to the full Monte-Carlo model.  This simplified model implementation is light enough for larger DNNs, and is described in Supplementary Materials.

\begin{figure*}[t]
\centering
\centerline{\epsfig{file=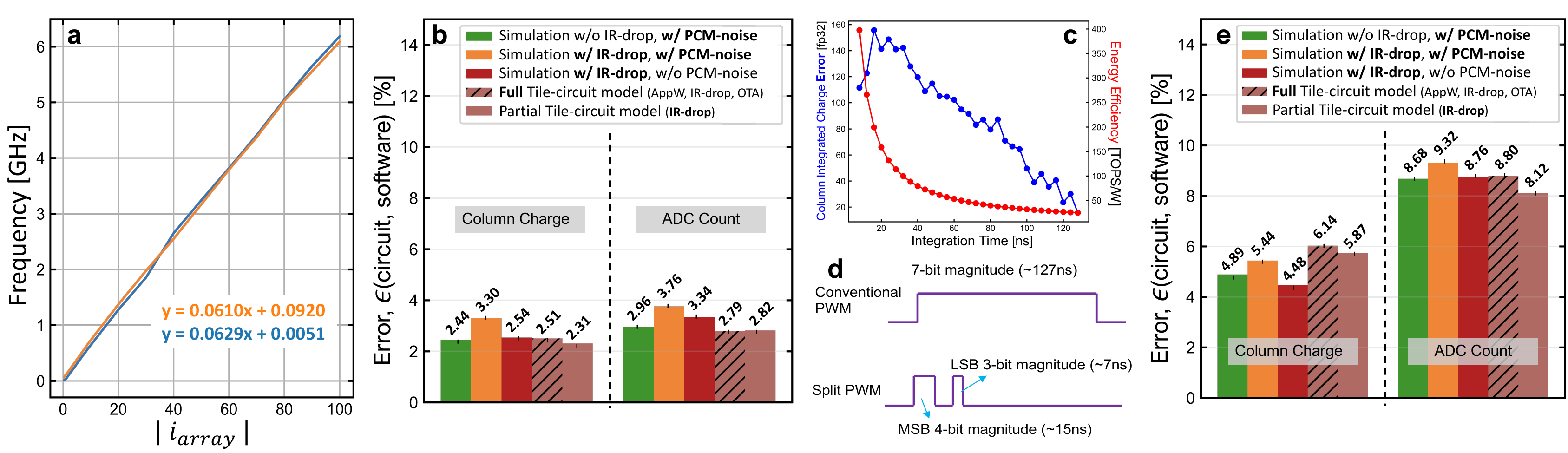, width= 1\columnwidth}}
\caption{\textbf{ADC transfer curves and python-based Tile-circuit model MAC results} (a) ADC transfer curve shows that, after appropriate calibration,  
the frequency of CCO pulses is highly linear as a function of incoming current.  (b) Error $\epsilon$ obtained with both full circuit-simulations and with our python-based Tile-circuit model, with respect to software MAC, at the integrated-current stage as well as the final digitized ADC output,
under different device and circuit non-ideality conditions with Conventional Pulse-Width Modulation (PWM) scheme.  A strong match is observed between the Tile-circuit model with all non-linearities (light brown solid bar) and full circuit simulations without PCM noise (reddish-brown solid bar), and the act of digitizing the continuous current (right-half vs.\ left-half of (b)) has only a modest effect on MAC error.  PCM noise is observed to contribute about the same amount of MAC error as IR-drop (compare green and reddish-brown bars); fortunately, IR-drop and PCM noise effects appear to add sub-linearly, since including both effects (orange bar) increases but does not double overall MAC error. 
Reducing the integration time -- either by (c) brute-force scaling the inputs into a shorter integration window, or (d) by switching from a Conventional PWM to Split PWM mode -- greatly improves the energy-efficiency, at the price of higher MAC error. 
(e) MAC error for Split PWM, under the same conditions as (b), shows that Split PWM increases error at the integrated current stage by 1.76$\times$, and at the digitized ADC output stage by 2.62$\times$, due to the effects of quantization error.  However, this is still smaller than the $\sim$7$\times$ increase observed in (c, dotted-line) for the equivalent integration-duration of 22ns while providing similar energy-efficiency benefits. 
}
\label{fig:simResults}
\vspace{-0.15in}
\end{figure*}

\section{ADC Behavioral Model }\label{adcModel}

Thus far, our analysis has focused on the instantaneous and integrated column current. 
However, this is only an intermediate stage -- the true determinant of MAC accuracy is the final digitized ADC output. 
Our simulation framework allows us to evaluate MAC accuracy at both of these stages within the operation, which allows us to independently characterize sources of error such as the ADC quantization effects inherent in digitizing the continuous column current. 

To incorporate the impact of the ADC digitization circuits, we apply the ADC transfer curve for oscillation frequency as a function of input current, as taken directly from full circuit simulations (Fig.~\ref{fig:simResults}a), to turn current over time into digitized counts. 
This transfer curve can be incorporated either as a simple linear fit, or as a piecewise-linear Look-Up Table (LUT). 
Using estimated/predicted currents from either full circuit simulations or our Tile-circuit model as input, we can use the ADC transfer function to produce a final digitized MAC output. 
While this study focuses on one particular Current-Controlled Oscillator ADC design, our general implementation approach can be used to evaluate and compare various alternative ADC designs.

With this understanding of the two checkpoints at which the MAC accuracy is evaluated and quantified, we can then revisit the error metrics referenced previously. 
Fig.~\ref{fig:simResults}b depicts the 2.54\% baseline error that was seen from circuit simulations (reddish-brown solid bar). 
The results of the python-based Tile-circuit model, where the goal is to match the error produced by the circuit simulation, show close correspondence, with an IR-drop-only model producing an error of 2.31\% (light-brown hashed bar), and one with all three non-idealities (apparent weight, IR-drop and OTA imperfections)
showing an error of 2.51\% (light-brown solid bar).  
The LUT version of the PCM noise-model from Section~\ref{sec:deviceModel} was integrated into the circuit simulation framework to generate MAC errors with PCM noise in Fig.~\ref{fig:simResults}c, showing that PCM read noise contributes to a similar level of MAC error as IR-drop (compare the green and the reddish-brown bars).  However, these noise sources add sublinearly, since modeling with both increases overall MAC error without doubling it (orange bars).

\subsection{Split Mode Pulse Width Modulator (PWM)}\label{sec:splitPWM}

When designing AIMC tiles, we aim to maintain accuracy and maximize energy efficiency. 
While modeling the accuracy impact has been the focus thus far, in some cases trading off accuracy for energy efficiency might be appropriate for a particular application. 
A key factor that determines energy efficiency is the time duration during which input activations are driven and ADC integration occurs. For the conventional PWM mode we have mostly discussed, this calls for 127ns of integration to cover the 7-bits of magnitude for a 8-bit signed input signal.  This does not include any additional time needed for control signals or design choices such as extra hold time introduced between PWM initialization and ADC activation. 
Minimizing the time during which circuit components need to remain active is a clear candidate for reducing power consumption. 
Fig.~\ref{fig:simResults}c demonstrates the accuracy and energy-efficiency tradeoff observed when crudely attempting to shorten the integration window by scaling down the input activation bit precision -- while there is dramatic increase in the energy-efficiency for shorter integration times, the MAC accuracy does degrade. 

Another approach to reduce integration time while maintaining input activation precision is the Split-mode PWM. 
For Conventional PWM, an activation magnitude is simply encoded into a voltage duration, as shown in Fig.~\ref{fig:simResults}d. 
In Split PWM mode, an activation is split into two voltage pulses: the first pulse encodes only the upper 4 MSB bits (of the activation magnitude), while the second pulse encodes the lower 3 LSB bits.  
During the MSB phase of the integration, a control signal in the ADC Counter causes every CCO pulse to be counted as 8 counter-ticks. 
In the LSB phase, this control signal is adjusted so that every CCO pulse is counted as a single counter-tick, returning to the conditions applicable to the Conventional PWM mode. 
Because every nanosecond within the MSB phase is amplified 8$\times$, the required total integration window is reduced to only 22ns (15ns for the MSB; 7ns for the LSB) to integrate activations with 7-bits of magnitude. 
This reduction leads to improved energy efficiency -- the Conventional PWM mode achieves an average performance of 26 TOPS/W, whereas Split PWM mode achieves an average performance of 106 TOPS/W. 

However, there is an accuracy trade-off, as can be observed by the increases in baseline MAC error shown in Fig.~\ref{fig:simResults}e.  
A primary source of additional error is ADC quantization -- any residual charge sitting on the active CCO capacitor at the end of any integration-phase represents array current that was integrated in the analog domain, but which failed to successfully trigger a counter-tick, and thus was not measured in the digital domain.  In the MSB phase of the Split PWM mode, the loss of this final CCO oscillation represents an error of up to 8 counter-ticks; while in the Conventional PWM mode or the LSB phase, the maximum error is only 1 counter-tick. (In general, the average error would be 4 counter-ticks and 0.5 counter-ticks, respectively.) 
Once again, comparison between the brownish-red and light-brown bars in Fig.~\ref{fig:simResults}e
shows that our Tile-circuit model accurately captures this additional source of MAC error, matching the predictions of full circuit simulations both in terms of column charge and ADC count.   As with the Conventional PWM mode, PCM-noise has a similar role to IR-drop  (compare the green and reddish-brown bars in Fig.~\ref{fig:simResults}e), and these two noise sources add sublinearly (orange bars).

\section{DNN Accuracy Simulations using AI Hardware ToolKit}\label{sec:aihwtk}

\begin{figure}[t]
\centering
\centerline{\epsfig{file=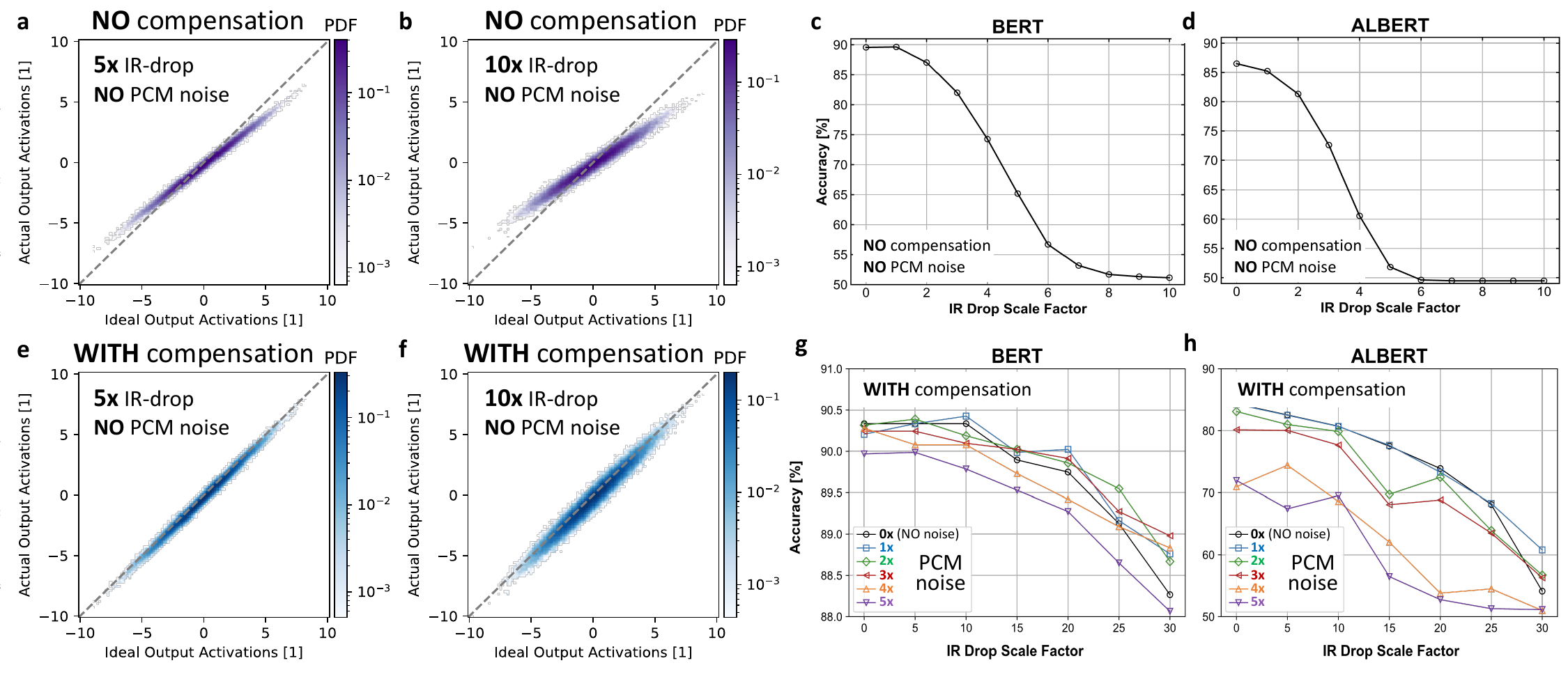, width=1\columnwidth}}
\caption{\textbf{BERT and ALBERT DNN-model accuracy on QNLI from IBM Analog Hardware Acceleration Kit:}  In the absence of compensation, (a)-(b) correlation plots between actual and ideal (desired) output activations -- obtained with AIHWKit for integration of a single 512$\times$512 linear layer under Conventional PWM excitation -- demonstrate that IR-drop has a strong de-correlating impact on output activations.
The sources of noise blurring the correlations include tile-parasitics, IR-drop at the specified relative wire-resistance level, and ADC quantization. 
(Neither PCM-noice nor apparent weight nor OTA imperfections are included yet).  
(c) While the BERT model prepared with Hardware Aware (HWA) training but inferenced without compensation shows some resilience, increases in wire resistance cause IR-drop to rapidly degrade DNN-model task accuracy towards random guessing (50\% accuracy).  Note that during the actual HWA training, the network is exposed only to simple random Gaussian noise.
(d) Similar trends are observed for the ALBERT models -- even modest IR-drop causes marked accuracy degradation. 
To further improve DNN-model resilience to IR-drop, a calibration procedure was developed that determines and applies compensation factors at each MAC output each during DNN inference with the QNLI dataset. 
(e)-(f) Correlation plots show that this compensation procedure does not markedly change the random noise but greatly improves the tracking between actual and ideal (desired) activations, for a single 512$\times$512 linear layer.  PCM-noise is still not yet included.
(g)-(h) Similarly, while
HWA training prepares the DNN-model to be resilient to various types of random noise (now including quantization noise and PCM-noise at various specified levels), inference with this compensation procedure exhibits much higher resilience to IR-drop, for both BERT and ALBERT models.  As previously observed~\cite{raschHWA:2023}, ALBERT is inherently more sensitive to noise than BERT.  Similar to the way that IR-drop is amplified by simply scaling the wire-resistance, PCM-noise is amplified by simply scaling up the magnitude of the particular Gaussian sigma that would otherwise be applied (based on batch-level average $t_{integration}$ as described in the Supplementary Materials).
} 
\label{fig:aihwkit}
\vspace{-0.15in}
\end{figure}

In this section, we describe the integration of the python-based Tile-circuit model and the PCM read noise-model into a PyTorch-based DNN framework that simulates full deep neural networks to demonstrate the impact of these low-level behaviors on end-to-end DNN inference. 
This simulator -- the IBM Analog Hardware Acceleration Kit (AIHWKit) -- is an open-source PyTorch-based toolkit that enables AIMC simulations for DNN training and inference~\cite{aihwkit_aicas, le2023using}. 
More details on AIHWKit can be found in the Methods section. 
The models used for this study are BERT Base and ALBERT Base V2, fine-tuned on the QNLI task from the GLUE dataset~\cite{wang_glue_2019}.

\subsection{AIHWKit Circuit and Device Model Implementation}\label{aihkwitCircuitModel}

As discussed in section \ref{sec:circuitModel}, column-wise IR-drop has the strongest impact on circuit-level MAC accuracy. 
We implemented the time-dependent IR-drop Tile-circuit model into AIHWKit, allowing the effective column-wire resistance (per Unit Cell) to be scaled from 0$\times$ (no IR-drop), to 1$\times$ (same column-wire resistance used in the circuit-simulation and tile-circuit results shown earlier, Fig.~\ref{fig:simResults}), and then to values as high as 10$\times$.  The AIHWKit implementation also includes the ADC response shown in Fig.~\ref{fig:simResults}a, providing an internal oscillation of approximately 6 GHz for an absolute current of 100$u$A, and the resulting quantization error effects. 

To implement device-level read noise non-ideality in AIHWKit, we use the most simplified (right most) noise-model in Fig.~\ref{fig:pcmModel}i, which samples an empirical fit of the standard deviation of noise as a function of expected conductance and mean batch read time. 
While the ADC quantization error is always implemented in inference, we explore the impact of IR-drop alone (in Fig.~\ref{fig:aihwkit}c-d) before including all three effects, including PCM read noise, in the Analog tile models in Fig.~\ref{fig:aihwkit}g-h.

\subsection{DNN Accuracy and Hardware-Aware Training}\label{aihkwitHWA}

Hardware-aware (HWA) training is commonly used to enhance DNN-model resilience to circuit and device non-idealities. 
During HWA training with AIHWKit, simple Gaussian noise is added to the DNN-model weights, which encourages the backpropagation algorithm to find a set of weights that offers low classification error on the training set as well as resilience to weight errors.
Using Conventional PWM excitation, Fig.~\ref{fig:aihwkit}c-d shows DNN inference results from BERT and ALBERT DNN-models after HWA-training.  During inference, AIHWKit includes our Tile-circuit integration effects including tile parasitics and quantization noise, but no PCM noise, and varying amounts of IR drop as parameterized by the relative column-wire resistance. 
The DNN-models exhibit some resilience to random noise when IR-noise is absent or minimal, but accuracy decreases quickly when IR-drop increases. 

To further improve DNN-model resilience to IR-drop, we implemented a calibration procedure prior to running DNN inference to compensate for the average signal loss due to IR-drop. 
The calibration procedure works as follows: we first take a set of input activation vectors -- for example, from a subset of the training dataset -- and compute the expected software MAC results without IR-drop. 
We then use the same input activations for the Analog tile models to compute the MAC outputs under IR-drop. 
Compensation factors are calculated from the two sets of MAC results, producing one shift and one scale factor for each output column in Analog tiles, which acts to take the average column activation in the presence of the given amount of IR-drop and scale it back to the average column activation exhibited by the original DNN-model in the absence of IR-drop.
Since these compensation factors can be combined with the existing affine-scaling factors, no additional DNN-model parameters need to be stored.  The impact of this calibration procedure is demonstrated on a single (512$\times$512) linear layer in Fig.~\ref{fig:aihwkit}a-b (inference without any compensation) and e-f (inference with compensation). 
A similar calibration procedure is also applied in experimental demonstrations and is also commonly used to compensate for hardware variations\cite{ares_nature}.

Fig.~\ref{fig:aihwkit}g-h demonstrate that a HWA trained DNN-model with IR-drop compensation significantly improves the ability of the HWA-trained DNN-model to remain resilient in the presence of IR-drop. 
Note that the exact analytical shapes developed in this paper were used only in post-training inference -- inference passes during training and validation used generic Gaussian distributions to keep training time tractable, yet the resulting DNN-models display significant robustness towards more exact representations of true hardware noise, even if these are non-Gaussian noise sources.  This is especially true for PCM read noise, as demonstrated by the various curves in Fig.~\ref{fig:aihwkit}g-h.

In the same way the HWA training can use simple Gaussian noise to prepare a DNN-model for exposure to non-Gaussian random noise, better DNN-model resilience to IR-drop should result if the DNN-model can be exposed during HWA-training to some sort of approximate yet computationally-tractable algorithm that adequately mimics the effects of IR-drop.
This was not implemented here due to the improved yet still significant compute complexity of the circuit and device models described here, but should be a goal of future work.

\section{Discussion}\label{sec:conclusions}

In this paper, we develop a methodology to model Analog tile circuits and devices that includes circuit simulations of IR-drop, transient analog signal responses, various circuit non-idealities and read noise. 
We analyzed the contributions of these non-idealities to MAC errors and developed a Tile-circuit model that can be integrated into DNN frameworks to evaluate inference accuracy of larger DNN models, such as a transformer. 
The analytical Tile-circuit model approach allows for proper understanding of the impact on DNN accuracy of various hardware imperfections and noise-sources, and the potential DNN-model resilience to these types of non-idealities. 
An experimentally calibrated device noise-model for the relevant integration-time regime of interest (1-200ns) allows for more precise analysis of hardware impact. 
Combining accuracy simulation from DNN frameworks and power performance estimations from circuit simulations, this methodology enables end-to-end AIMC design optimization for target DNN applications such as Large Language Models (LLMs).

In terms of hardware-aware (HWA) training, while simulating the exact inference path during DNN training can be costly computationally, generic noise-aware training approaches show some resilience. 
Post training calibration techniques can be applied, typically using additional validation passes, to improve AIMC inference accuracy without extensive DNN-model fine-tuning~\cite{lammie2024improving, huang2024opposing}. 
As a result, accurate AIMC inference models can take on slightly more compute complexity and still serve as a means for accuracy verification. 

Some design optimizations for the Analog tile were discussed and simulated, including the use of level-shifter to avoid low current MAC errors and the use of Split PWM mode to increase MAC energy efficiency at a cost of reduced MAC accuracy. 
While these optimizations are only a few examples of possible Analog tile improvements, we demonstrate a methodology to quantify the benefits of these improvements via MAC errors and energy efficiency. 
Additional PWM modes beyond Conventional and Split PWM can also be explored in future work.

\section{Methods}\label{sec:methods}

\subsection{Average Energy Efficiency}\label{energy_efficiency}

Energy efficiency, measured in [TOPS/W], was estimated for a MAC operation by measuring the current through all voltage sources in the simulation: peripheral circuitry, unit cell read voltage and transistors, and the per-column ADC. 
Importantly, a MAC operation on a single column consists of a total of 512x2 operations -- 512 Unit Cells in a column, each with two conductance pairs. 
The current was measured for the initial control circuitry timing, CCO hold time and true MAC integration window. 
The final energy efficiency reported is the average across 200 simulation points, each of which utilizes Gaussian-distributed activation/weight inputs vectors. 

\subsection{AIHWKit Functionality}

The core of the toolkit is based on objects referred to as Tiles. 
The weights of a given neural network are mapped onto one (or more) Tile(s); the toolkit allows for customization of the mapping, including specifying a Tile size (input and output dimensions), padding and various input and output scaling factors, among other parameters. 
The primary function of this Tile mimics the function of the analog tile described in the circuit simulation framework from Section \ref{sec:overview}: to perform a forward inference pass, which consists of computing various MAC operations. 
Similarly, the toolkit allows for various customizations of the forward inference, including setting input and output resolutions, clipping, bounding and the addition of various noises with tunable characteristics. 

The extensive parameter space of AIHWKit is the driving capability that allows for thorough understanding of AIMC for deep neural network accuracy. 
While the toolkit has various parameters that model the hardware on which the DNNs are mapped, trained and inferenced, there continues to be trade-offs between complexity of circuit and device models and the speed of DNN simulations. 
The incorporation of circuit-level and PCM noise device-level models in this paper resulted in a $\sim$10$\times$ increase in DNN inference time, even after optimizing the python-based models with C++ libraries. 
Further speed improvements on the accurate low-level models or higher level approximations of these models would be needed to simulate larger models and to include more of these effects within the hardware-aware training loop.

\section{References}
\bibliography{references}

\section{Acknowledgment}

We thank V. Mukherjee, W. Wilcke, S. Narayan, S. Munetoh, S. Yamamichi, M. Furmanek, C. Osborn, J. Burns, R. Divakaruni, M. Khare and the IBM Research AI Hardware Center for management and logistical support.

\clearpage
\setcounter{figure}{0}
\renewcommand{\figurename}{\textbf{Supplementary Fig.}}
\renewcommand{\thefigure}{\textbf{S\arabic{figure}}}

\section{Supplementary Materials}

\subsection{Analog Tile Design and Analysis}\label{circuitAnalysis}

To model the analog MAC operations with reasonable computational resources and to assess the accuracy impact of analog non-idealities, we simulate a single column (a 512$\times$1 Tile).
The structure has the necessary flexibility to model effects from neighboring columns and rows, and allows us to compare the circuit simulation results against an ideal MAC.

\begin{figure}[b]
\centering    
\centerline{\epsfig{file=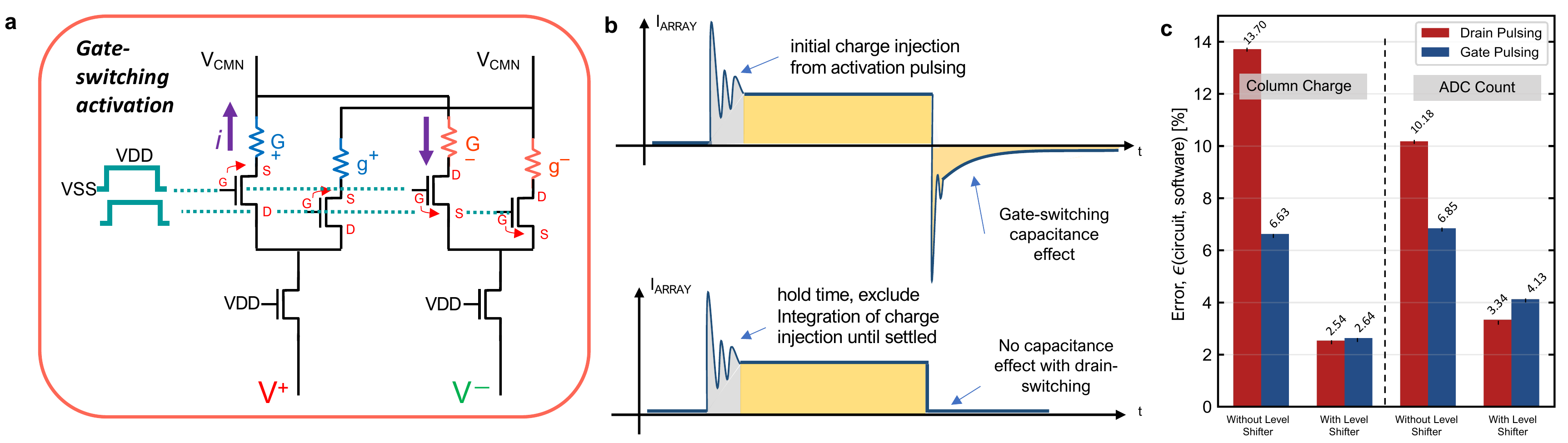, width=0.9\columnwidth}}
\caption{\textbf{Circuit Design} (a) For Gate-Switching activations, inputs are driven to the gate terminals of select transistors, while the drain/source terminals are held at a constant $V_{cmn}$ and $V^{+}$/$V^{-}$, dependent on the sign of the activation. 
(b) Both activation schemes incur an initial charge injection, which can be waited out by introducing a ``hold time'' between initializing the PWM outputs and enabling the CCO and PWM counters, thus avoiding any integration of unwanted charge.  However, only the Gate-Switching scheme incurs an additional capacitance effect --  unwanted charge being integrated as floating nodes settle after a transistor is turned OFF. 
The illustration shows a `peak' capacitance effect where all these effects pile up at the end of an integration window, by mandating that all devices turn off together at the end of the integration interval.  However, even when transistors turn off within the integration window, while the effects may not stand out on a plot of current, each unit-cell is inherently contributing undesired current even after its intended shut-off.  
(c) Error between circuit and software simulations under various simulation conditions. 
Drain-Switching activation scheme, combined with level-shifter, results in lower error than Gate-Activation.}
\label{fig:supColSims}
\vspace{-0.15in}
\end{figure}

\subsubsection{Input Gate-Switching vs Drain-Switching and Hold Time}\label{input_switching}

We compare two designs to drive input activation signals to the Unit Cell (UC). 
Fig.~\ref{fig:simulationFramework}c illustrates the first option: driving activation pulses into the drain terminals of select transistors (Drain-Switching Activation); 
Fig.~\ref{fig:supColSims}a demonstrates the second: driving the gate terminals of select transistors (Gate-Switching Activation). 
In either mode, the transistors which are not being driven by the activation pulse are held LOW for the duration of the integration time. 

In both switching modes, the circuitry is vulnerable to instability during the start of the integration window. 
As a MAC operation begins, 512 activations are simultaneously applied to all 512 UCs in a column, taking them from a previously OFF state to an ON state. 
This rapid, simultaneous activation of all UCs causes a charge injection into the column, which settles after some time. 
In order to avoid integrating current from this unwanted effect, a hold time is implemented. 
During this programmable hold time, all usual control circuitry is enabled for normal operation, with exception of the CCO; after the hold time is satisfied, control signals enable the CCO and the current is then officially integrated for the desired integration period, and counted towards the final MAC result. 

An effect not shared by both modes is the gate-switching capacitance effect. 
In the gate-switching activation mode, the activation is pulsed to the gate terminal of the select transistor to turn it ON; this allows for a voltage potential to be created across the terminals of the PCM device -- with one end being tied consistently to $V_{cmn}$ while the other is tied to $V^{+}$/$V^{-}$ (depending on the sign of the activation). 
After the designated activation duration, the gate of the transistor is set to $V_{SS}$, driving the transistor OFF. 
At this point, the node of the PCM device that was previously being driven to $V^{+}$/$V^{-}$ is left to settle to $V_{cmn}$. 
During this settling time, some amount of current is still being generated, which contributes unwanted charge integration to the MAC. 
Fig.~\ref{fig:supColSims}b illustrates an over-simplified version of this effect, where all activations end at the exact same time -- meaning all 512 UCs are simultaneously experiencing this capacitance effect and settling to an OFF state. 
Realistically, activations will end throughout an integration period, causing charge contributions from this capacitance effect to be staggered across the entire integration window (and thus not as easily visualized as in this simplified conceptual diagram), though they will have a cumulative effect similar to that illustrated. 

This effect is namely a gate-switching capacitance effect; in the drain-switching mode, this effect is avoided by forcing OFF state activations to $V_{cmn}$, rather than letting them settle independently. 
The trade-off of this mode, however, manifests in a different effect: $V_{cmn}$ mismatch between the two nodes. 
This effect can be accurately captured by the IR-drop Tile-circuit model, because turned-off rows are part of the Thevenin equivalent calculation. 
Drain switching has better energy efficiency due to the smaller voltage swings and lack of $CV^2$ switching. 
Unless otherwise stated, drain-switching is the mode of operation that is implemented in the analysis in this paper.

The average energy efficiency of the two modes is comparable: gate-switching achieves 35.15 TOPS/W, and drain-switching achieves a higher 37.75 TOPS/W, for Conventional PWM activation.

\subsubsection{Current level-shifter for Low Current Operation}\label{levelShifter}

While any fixed shift and scaling factor can be calibrated on-chip at the ADC output, we observe large MAC errors when expected SW MAC is small. 
This is because under low array currents, the regulator FETs in the ADC mirror fall into the sub-threshold regime, leading to disrupted voltage regulation. 
Therefore, we introduce the level-shifter in Fig.~\ref{fig:simulationFramework}b to address this. 
The level-shifter circuitry injects current and prevents those regulator FETs from falling into sub-threshold. 
The strength of the level-shifter bias-current is tuned via a control signal. 
In this design, the effects of the level-shifter are prominent when the current generated by the array is in the few uA regime, and become negligible for larger currents. 
As such, the level-shifter helps reduce MAC error in lower current ranges, while not negatively impacting/biasing MAC operation in higher ranges. 
Fig.~\ref{fig:supColSims}c captures the impact the level shift circuitry has on reducing MAC error. 
Unless otherwise stated, the level-shifter is assumed to be ON in the analysis in this paper.

The error correction benefits from the level-shifter come with a small energy efficiency cost. 
The average energy efficiency is 25.50 TOPS/W when the level-shifter bias is tuned to its maximum range. 
The strength of the level-shifter can be tuned to explore intermediate accuracy-energy trade-offs.

\clearpage
\subsection{PCM Device Experiment Setup}\label{pcmExperiment}

\begin{figure}[b]
\centering    
\centerline{\epsfig{file=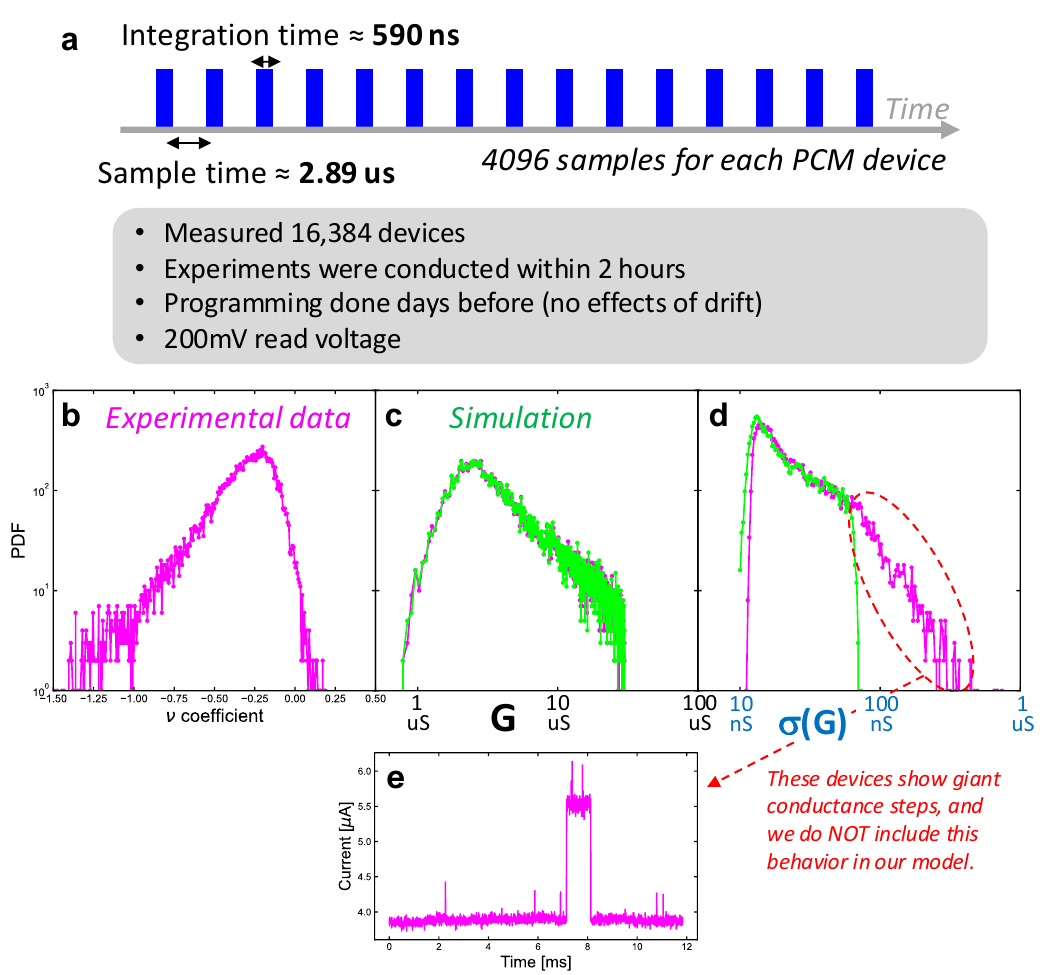, width=0.9\columnwidth}}
\caption{\textbf{PCM Data measurement and selection} (a) Experiment conditions for measuring a 512x32 array of PCM devices. 
(b) Drift coefficients of experimental data (c) Experiment and simulated conductance distributions demonstrate high correspondence. 
(d)-(e) Experimental PCM conductance traces demonstrate a population of devices with large conductance steps, indicative of Random Telegraph Noise (RTN) --  device behavior that is not included in the noise-model.
}
\label{fig:pcmDataCollection}
\vspace{-0.15in}
\end{figure}

To experimentally characterize PCM device read noise, we setup a testbench with a 512x32 array of Phase-Change Memory (PCM) devices, programmed to reach a distribution of conductance values and a read with a 200mV read voltage, as shown in Fig.~\ref{fig:pcmDataCollection}a-c. 
The current generated by each device was sampled and integrated for 590ns, and each of 4096 total samples was taken with a sampling time of 2.89$u$s. 
Importantly, the PCM devices were programmed long before the measurements were taken to minimize the effects of PCM conductance drift. 
We chose not to model the effect of random telegraph noise (RTN) in this case as it is an effect that has impact only over much longer time-scales.

\clearpage
\subsection{PCM Read Noise Modeling}

\begin{figure}[b]
\centering    
\centerline{\epsfig{file=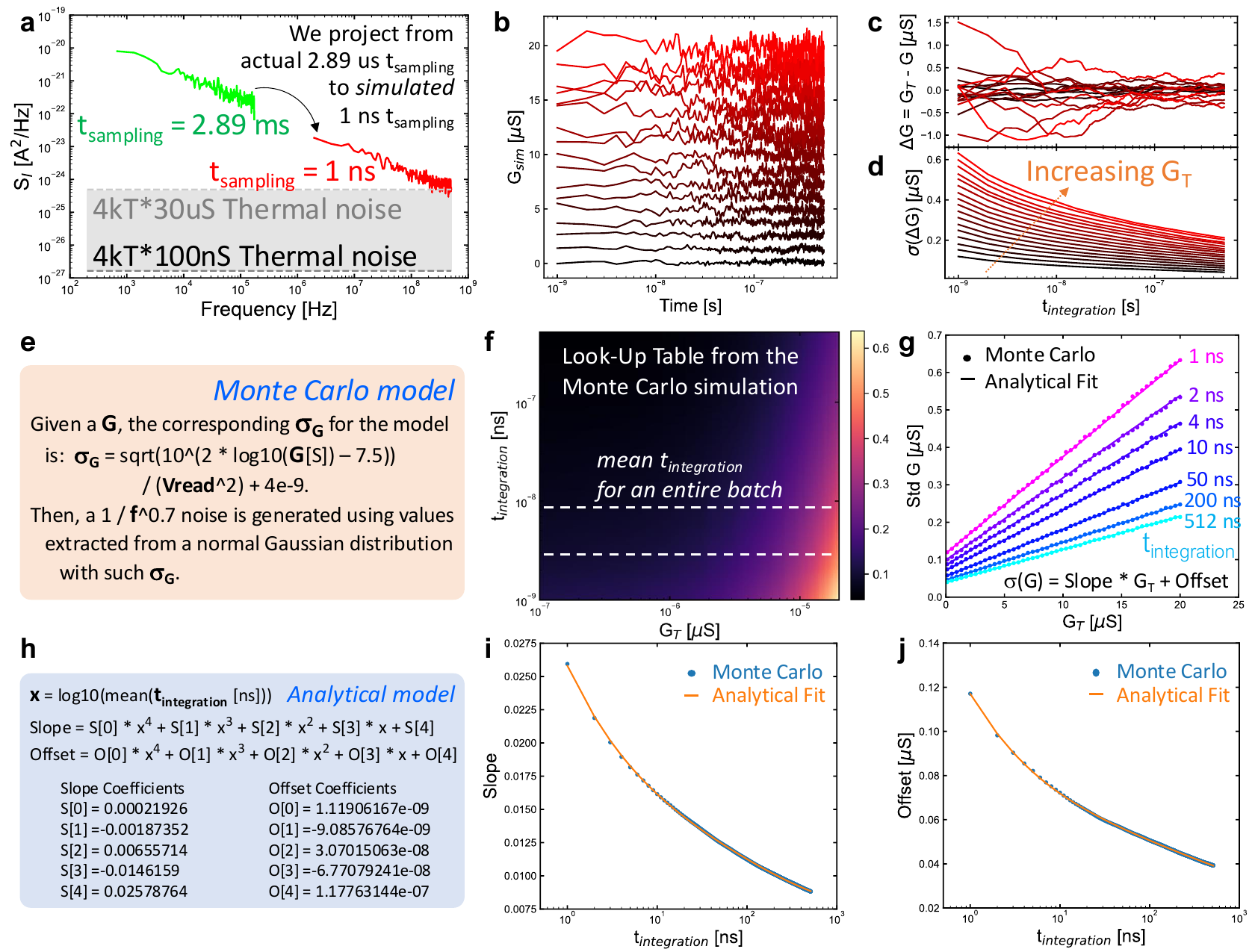, width=0.9\columnwidth}}
\caption{\textbf{PCM Monte Carlo, LUT and Analytical noise-model} (a) Experimental power spectrum data is projected from the actual sampling time down to simulated sampling time to capture noise behavior in relevant nanosecond time regime. 
(b) Simulated PCM conductance behavior with $1/f$ noise from nanosecond projected power spectrum. 
(c) Instantaneous conductance deviation from the target conductance at a given integration time. 
(d) By simulating several thousand PCM conductance traces, for a given target conductance, a sigma value (representing the broadening of the traces) can be extracted for a given integration time. 
(e) Monte Carlo noise-model fitted from experimental data and simulations. 
(f) A variation of the PCM device noise-model to further reduce computational complexity does not extract a noise sigma per conductance/activation pair, but rather a noise sigma dependent on the mean activation magnitude of an entire batch. 
(g)-(j) An analytical abstraction of the per-batch, integration time dependent LUT developed to facilitate efficient integration into higher-level simulation frameworks.
}
\label{fig:pcmModeling}
\vspace{-0.15in}
\end{figure}

Based on PCM experimental data, we build a Monte Carlo noise-model to generate simulated PCM device data, that accurately tracks the experimental data. 
We first project the experimental power spectrum down to shorter sampling time in Fig.~\ref{fig:pcmModeling}a. 
Note that the power spectrum still lies above thermal noise. 
We then simulated individual conductance traces over time to generate statistical data in Fig.~\ref{fig:pcmModeling}b-d. 
The final Monte Carlo noise-model is described in Fig.~\ref{fig:pcmModeling}e.
 
Fig.~\ref{fig:pcmModeling}f visualizes one such noise-model as a heatmap/lookup table (LUT), where the PCM device noise is a function of both the target programmed conductance and the integration time for the respective device. 
Fig.~\ref{fig:pcmModeling}g shows slices from Fig.~\ref{fig:pcmModeling}f for different mean $t_{integration}$ and their corresponding analytical fit. 
The fit coefficients are shown in Fig.~\ref{fig:pcmModeling}h-j. 

\end{document}